\documentstyle[prl,aps]{revtex}
\begin{document}
\draft
\title{Puzzlement about thermal redshift}
\author{Vesselin I. Dimitrov}
\address{Faculty of Physics, Sofia University, BG-1164 Sofia, Bulgaria}
\date{\today }
\maketitle

\begin{abstract}
Discussed is the classical theoretical description of the experimentally
established thermal redshift of spectral lines. Straightforward calculation
of the observable spectrum from a canonical ensamble of monochromatic
radiators yileds overall blueshift rather than redshift. It is concluded
that the customary explanation of the thermal redshift as a second order
Doppler effect does not bear closer examination, and that in fact, the the
phenomenon ''thermal redshift'' is not yet fully uderstood in classical
terms.
\end{abstract}


\paragraph*{Introducion}

Two years after the first paper of M\"ossbauer on recoilless gamma resonance
fluorescence, Pound and Rebka observe a dependence of the effect on the
temperature\cite{ref1}, which they interpret as a second order relativistic
time dilation. Shortly after, Josephson advances a quatum--mechanical
explanation of the same phenomenon\cite{ref2}, pointing out that the quantum
and the classical pictures, although qualitatively very different, yield the
same quanitative result for the temperature dependence of the observed gamma
line's redshift. Later on this quantitative equivalence is callenged\cite
{ref3}, but the provoked discussion\cite{ref4,ref5} shows this to be based
on a misunderstanding. Since then the issue of the thermal redshift of
spectral lines is considered settled and becomes more or less a textbook
material.

It is the purpose of this letter to demonstrate, by an explicit calculation
of the spectrum, that the theoretical interpretation of the observed thermal
redshift is in fact in a very unsatisfactory state. This claim may seem
surprising at the end of the century marked with undisputably great
achievments in all branches of physics, but one has to remember that quantum
mechanics has occured at the beginning of the century exactly as an attempt
to better understand the emission and scattering of radiation by charged
systems -- during the years that followed, it has proven extremely useful in
its ability to predict the outcomes of all kind of experiments, but on
conceptual level confusion has entered the field and has never left it since%
\cite{ref7}.

Thermal redshift is most conveniently discussed (and is only observed) in
gamma spectroscopy. M\"ossbauer effect and associated redshifts of the
spectral lines have a special r${\rm {\hat o}}$le in the quantum--classical
controversy, because of the striking dissimilarity of the corresponding
descriptions. Within the quantum realm, a system radiating a photon
experiences recoil and the energy of the transition is distributed between
the radiator and the emitted photon in compliance with momentum and energy
conservation. In this context, the thermal redshift is caused by the
difference between the masses of the radiator before and after the emission
of the photon\cite{ref1,ref2}; and the M\"ossbauer effect occurs when the
emitter is a part of a solid and the recoil is experienced by the whole
lattice, thus being of negligible magnitude. Alternatively, Shapiro\cite
{ref13} has advanced a purely classical explanation of the above, based on
the frequency/phase modulation of the emitted radiation due to the motion of
its source. The quantitative predictions of the two approaches are basically
the same\footnote{%
The classical approach violates the Lipkin sum rule as it predicts satelite
lines with both lower and higher frequency. To the knowledge of the author,
whether or not such lines exist has never been checked experimentally, due
to the exceedingly small amplitude of the shifted spectrum.}. Now, when for
the same phenomenon more than one different ''correct'' descriptions exist,
it is a clear indication that, probably, we do not understand it at all. The
concept of photon, on which the quantum description is based, when taken
literally, has proven during the years to be an efficient generator of
confusion\cite{photon1,photon2}. Therefore, in this letter we explore the
classical approach to the thermal redshift, and by deriving the form of the
observable spectrum of a canonical ensamble of monochromatic radiators, we
demonstrate that here, too, a problem is encountered rather than a solution.

\paragraph*{Spectrum from canonical ensamble of radiators}

Let us consider an ensamble of identical massive particles, each of them
carrying a charged harmonic oscillator with a proper frequency $\omega _0$.
Let the paricles be constrained to move within a finite volume $V$, but
otherwise free and noninteracting with each other. Assume also that the
above ensamble is in equlibrium at temperature $T$ for the translational
degrees of freedom, and that the oscillators, powered by some internal
sources of energy, radiate steadily electromagnetic waves in the surrounding
space. Such a system models e.g. a container with an ideal gas of a
gamma-radioactive nuclide, observed during time small compared to its decay
half--time. It is a well established experimental fact that the radiation
spectrum of such system, instead of a sharp line at frequency $\omega _0$,
exhibits a line of finite width and central frequency redshifted with
respect to $\omega _0$. When the temperature $T$ is increased, so are the
observable line width and the redshift.

For our purposes we shall need to consider two reference frames $K$ (the
laboratory frame in which the observer is at rest) and $K^{\prime }$ (the
rest frame of an arbitrary member of the ensamble), connected by a Lorentz
transformation with dimensionless velocity $\bbox{\beta }$. Introducing the
four--acceleration

\[
a^\mu =\frac{du^\mu }{d\tau };\qquad a^\mu u_\mu =0
\]
where $u$ is the four--velocity and $\tau $ is the proper time, we have in
components

\begin{equation}
a^{\prime 0}=\frac{c\dot \beta ^2}{2(1-\beta ^2)^2}\qquad {\bf a}^{\prime }=%
\frac c{1-\beta ^2}\dot{\bbox{\beta }}+a^{\prime 0}\bbox{\beta }
\label{Eq1}
\end{equation}
from where we deduce the time derivative of the velocity, with $\gamma
=(1-\beta ^2)^{-1/2}$

\[
\dot{\bbox{\beta }}=c^{-1}(1-\beta ^2)\left[ {\bf a}^{\prime }-\frac \gamma
{1+\gamma }({\bf a}^{\prime }\cdot \bbox{\beta })\bbox{\beta}\right]
\]
Plugging this into the well--known formula for the intensity distribution of
the emmitted electromagnetic waves\cite{ref8} yields, after some algebra,

\begin{equation}
\frac{dI(t^{\prime })}{d\Omega }=\frac{e^2}{4\pi c^3}\frac{(1-\beta ^2)^2}{%
(1-\bbox{\beta }\cdot {\bf n})^4}\left[ a^{\prime 2}-({\bf a}^{\prime }{\bf \cdot
n}^{\prime })^2\right] _{ret}  \eqnum{2a}  \label{Eq2a}
\end{equation}
where ${\bf n}$ is an unit vector in the direction of the radiator. Having
in mind identical isotropic radiators, we have for the energy radiated in
unit time

\begin{equation}
\frac{d^2E(t^{\prime })}{d\Omega }=\frac{dI(t^{\prime })}{d\Omega }%
dt^{\prime }=\frac{dI(t^{\prime })}{d\Omega }\frac{dt^{\prime }}{dt}%
dt=\,I_0^{\prime }\frac{(1-\beta ^2)^2}{(1-\bbox{\beta }\cdot {\bf n})^3}dt
\eqnum{2b}  \label{Eq2b}
\end{equation}
where $\,I_0^{\prime }=\frac{e^2}{8\pi c^3}\overline{a^{\prime 2}}$ is the
average intensity of the radiated waves, identical for all radiators.

Assuming monochromatic sources with $S^{\prime }(\omega ^{\prime
})=I_0^{\prime }\delta (\omega ^{\prime }-\omega _0)$, we would have an
observable spectrum

\begin{equation}
S(\omega )=\,I_0^{\prime }<\frac{(1-\beta ^2)^2}{(1-\beta \cos \theta )^3}%
\delta (\omega -\omega _0\frac{\sqrt{1-\beta ^2}}{1-\beta \cos \theta })>
\eqnum{3}  \label{Eq3}
\end{equation}
where $\theta $ is the angle between $\bbox{\beta }$ and ${\bf n}$, and the
averaging is performed over the distribution of the source's velocity

Before proceeding with the averaging, Eq.(3) needs a comment. In the current
literature on the shapes of gamma lines, for reasons that remain obscure for
the present author, the factor in front of the delta--function is always
absent. In order to obtain the conventional thermal redshift, one needs a
factor of $(1-\beta \cos \theta )$, which reflects the retardation in the
propagation of the waves as indicated in Eq.(2b). Therefore, what is usually
missed, is exactly the factor $\frac{(1-\beta ^2)^2}{(1-\beta \cos \theta )^4%
}$, comming from the transformation of the wave intensity. In quantum terms,
this would imply that one photon is registered from every member of the
ensamble -- an assumption that completely disregards the abberation and the
changes in the rates of photon emission due to the Lorentz transformations%
\cite{abe}.\footnote{%
Curiously, the intensity factor has been properly considered by Sir Arthur
Compton in an attempt for a classical explanation of the Compton scattering%
\cite{compton}, but has been disregarded ever since.}

Since the sources are in equilibrium at temperature $T$ , $\cos \theta $ is
uniformly distributed in $[-1,1]$ and $\beta $ is distributed according to
the properly normalized J\"uttner's law\cite{ref9}

\[
P(\beta )=\frac \alpha {K_2(\alpha )}\,\frac{\beta ^2}{(1-\beta ^2)^{5/2}}%
\exp [-\frac \alpha {\sqrt{1-\beta ^2}}]
\]
with $\alpha =mc^2/kT$ and $K_2$ being the modified Bessel function of the
second kind. The averaging of Eq.(3) over the directions gives

\[
S(\omega )=I_0^{\prime }\frac \omega {\omega _0^2}\frac \alpha {2K_2(\alpha )%
}\int_{b_1}^{b_2}d\beta \beta \frac{\exp [-\frac \alpha {\sqrt{1-\beta ^2}}]%
}{(1-\beta ^2)^{3/2}}
\]
The integration limits above are determined from the requirement

\[
\omega \in \left[ \omega _0\sqrt{\frac{1-\beta }{1+\beta }},\omega _0\sqrt{%
\frac{1+\beta }{1-\beta }}\right]
\]
which can easily be shown to result in $b_1=\left[ (\omega _0^2-\omega
^2)/(\omega _0^2+\omega ^2)\right] ^2,\;b_2=1$. Performing the integration,
we obtain the observable spectrum

\begin{equation}
S(\omega )=\frac{I_0^{\prime }}{2K_2(\alpha )}\frac \omega {\omega _0^2}\exp
\left[ -\frac \alpha 2\left( \frac \omega {\omega _0}+\frac{\omega _0}\omega
\right) \right]  \eqnum{4}  \label{Eq4}
\end{equation}

The above formula, which is the main result of this work, describes an
asymmetric line centered at frequency

\[
\omega _0^{\prime }=\omega _0\left( \alpha ^{-1}+\sqrt{1+\alpha ^{-2}}%
\right)
\]
which is always higher than the proper frequency of the radiators $\omega _0$%
; thus special theory of relativity predicts blueshift rather than redshift
(see Fig.1). For low temperatures we have $\alpha \gg 1$, and the predicted
blueshift is proportional to the temperature,

\[
\frac{\Delta \omega }{\omega _0}=\alpha ^{-1}+O(\alpha ^{-2})=\frac{kT}{mc^2}%
+O(\alpha ^{-2}),
\]
a property shared with the blackbody radiation. The width of the spectrum
can be estimated from

\[
\overline{\omega ^2}-\overline{\omega }^2=\omega _0^2\left[ \frac{K_4(\alpha
)}{K_2(\alpha )}-\left( \frac{K_3(\alpha )}{K_2(\alpha )}\right) ^2\right]
=\omega _0^2\left[ \frac 72\alpha ^{-1}+O(\alpha ^{-2})\right]
\]
hence the witdth is proportional to $T^{1/2}$. When the temperature is let
to go up to infinity $(\alpha \rightarrow 0)$, the observed spectrum assumes
the Lorentz--invariant form

\[
S_\infty (\omega )=\left( \frac \alpha {2\omega _0}\right) ^2\omega
+O(\alpha ^3)
\]
again in accord with the behaviour of the blackbody radiation.

\paragraph*{Discussion}

Having derived Eq.(\ref{Eq4}), we find ourselves in an ambiguous situation.
On the one hand, the calculated observable spectrum shows features that are
generally expected on physical grounds, and by the analogy with the
behaviour of the blackbody radiation. On the other hand, no matter how
plausible the above arguments may be, experiment shows otherwise -- with
increased temperature centers of spectral lines are shifted towards lower
frequencies, period. One may argue that the purely classical treatment is
not adequate for the problem at hand; however, current practice in quantum
electrodynamics and quantum optics indicates that classical treatment cannot
be {\em that }wrong as to produce the {\em oposite} effect\cite{vleck}.

One option for reconciliation of the theoretical and experimental results
can be looked for in the nature of the registration process at the
observer's detector. If the frequency could be registered by a direct
counting of the waves arriving at the observer in one second (as can be done
e.g. for radiowaves), the intensity of the radiation would no longer be
relevant, and the observable frequency distribution would no longer be given
by Eq.(3), but rather by

\[
P(\omega )=\,<(1-\beta \cos \theta )\,\delta (\omega -\omega _0\frac{\sqrt{%
1-\beta ^2}}{1-\beta \cos \theta })>\,=\frac 1{2K_2(\alpha )}\frac{\omega
_0^2}{\omega ^3}\exp \left[ -\frac \alpha 2\left( \frac \omega {\omega _0}+%
\frac{\omega _0}\omega \right) \right]
\]
and the average observable frequency would be

\[
\overline{\omega }=\int_0^\infty d\omega \omega P(\omega )=\omega _0\frac{%
K_1(\alpha )}{K_2(\alpha )}=\omega _0<\sqrt{1-\beta ^2}>
\]
which is the customary expression of the thermal redshift.

In typical M\"ossbauer experiment, when direct frequency measurement is
clearly impossible, the setup includes a resonance detector attached to a
device that drives it with a prescribed velocity, thus Doppler--scanning the
profile of the gamma line. Alternatively, a resonance absorber can be placed
between the source and the detector and driven to attenuate various
frequency components. The counting rate is proportional to the intensity of
the waves of the corresponding frequency. It is not difficult to realize,
however, that, due to the motion of the detector (absorber), the different
frequency components of the registered line will be weighted with relative
weights $(\omega _d/\omega )^4$ where $\omega _d$ is the central frequency
of the detector's line, and thus the average registered frequency will
become $\omega _0<\sqrt{1-\beta ^2}>$ as above. If this explanation is to be
correct, scanning the gamma line by a method not based on Doppler effect
(e.g. by diffraction grating of based on Zeeman's effect) should reveal
thermal blueshifts instead of redshifts. Unfortunately, in frequency ranges
where such methods are feasible, the thermal redshift is much too small to
be observed. Of course, there yet remains the conceptual difficulty that in
the quantum description the redshift occurs at the source whereas in the
above classical description it should be due to the detector's motion, thus
no chance is left for constructing a limiting situation where the two
descriptions coincide.

Another interesting possibility to be considered is that the experimental
redshifts may occcur due to suitable correlations between different parts of
an extended source. That such effects do exists at all for electromagnetic
and sound waves has been realized theoretically\cite{wolf,ref10} and
verified experimentally\cite{ref11,ref12} only recently, although the
analogous phenomenon for neutron scattering is known for long\cite{hove}.
Classical electrodynamics is known to become inconsistent for distances of
the order and smaller than the classical electron radius. Hence, considering
truly point--like sources of radiation may not be correct. Allowing the
sources to have structure, one has to account for the possible effect of
this structure on the observable far--zone spectrum of the radiation, and it
has been demonstrated that correlations within the source are capable of
producing Doppler--like redshifts and blueshifts of the spectral lines\cite
{ref11,ref12}. Moreover, a rigorous quantum--mechanical treatment of
gamma--line redshifts, not based on a corpuscular picture of the photon, but
closely analogous to that of neutrons\cite{sing} is capable of predicting
spectral shifts clearly associated with source's correlations. Thus, if this
turns out to be the correct explanation, it would be possible to construct
meaningful classical limit where the classical and the quantum descriptions
merge into one another.

Whatever the truth may be, we have to conclude that the phenomenon of the
thermal redshift is not yet fully understood. Clarifying the origin of the
descrepancy between the observations and the theoretical expectations will
certainly bring about deeper understanding of the processes of the
interaction between matter and radiation, and perhaps, of the relationship
between classical and quantum physics.

\paragraph*{Acknowledgments}

The author thanks the foundation ''Bulgarian Science and Culture'' for its
generous support, and Dr. Derek Abbott for his interest in this work.

\begin{figure}[tbp]
\caption{Spectral line profiles as obtained from Eq.(4): The parameter $%
\alpha $ has value 10 for the lowest curve, 20 for the middle, and 30 for
the highest one.}
\end{figure}

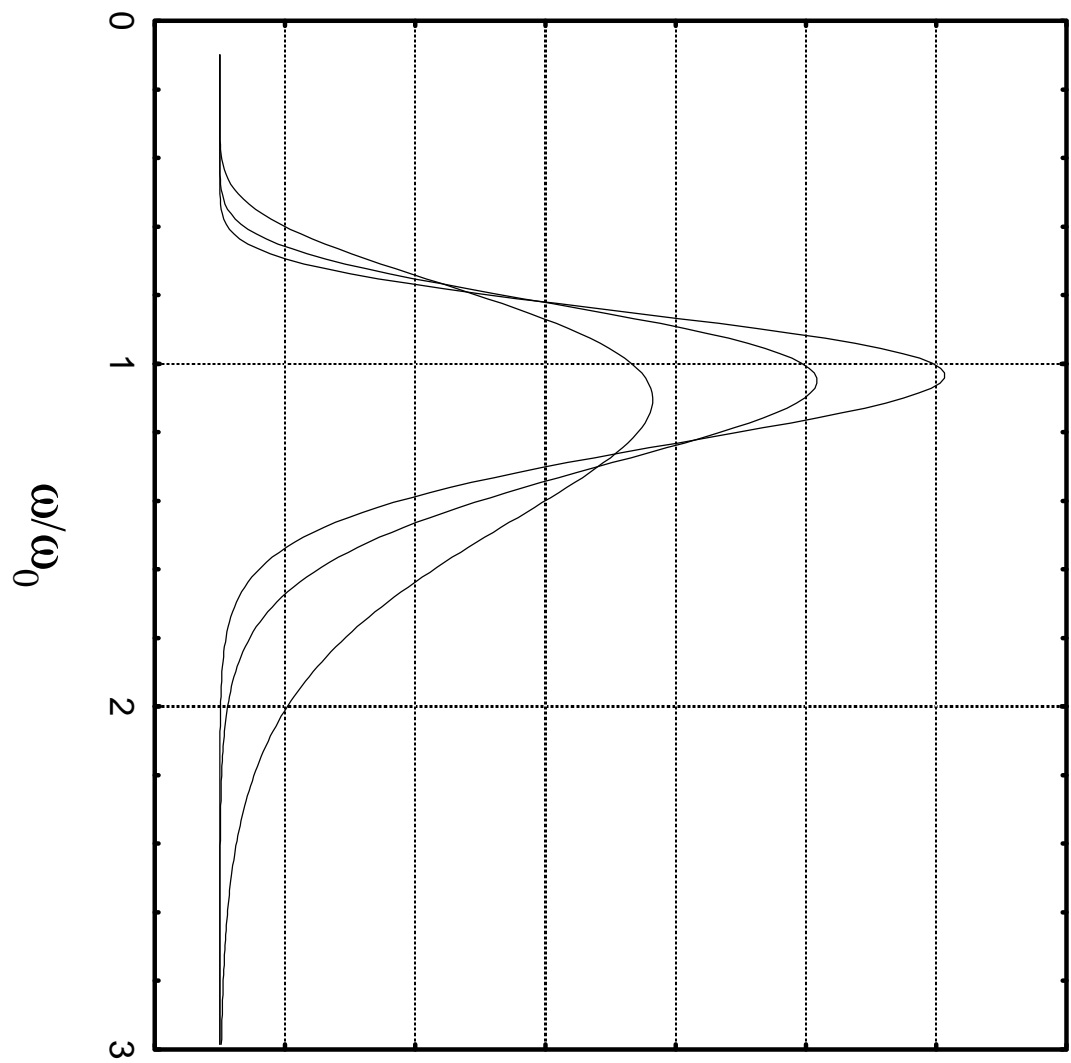
\end{document}